\documentclass[english,journal=jacsat,manuscript=article,etalmode=truncate,maxauthors=0,layout=twocolumn]{achemso}
\setkeys{acs}{etalmode=truncate,maxauthors=0,articletitle=true,email=true}

\usepackage{amsmath}             
\usepackage{mathtools}             
\usepackage{amssymb}             
\usepackage{wasysym}             
\usepackage{color}               
\usepackage{bm}
\usepackage{setspace}            
\usepackage{graphicx}            
\usepackage{diagbox}
\usepackage{caption}
\usepackage{subcaption}

\usepackage{wrapfig}             
\usepackage[dvipsnames]{xcolor}  
\usepackage{array,booktabs,multirow}      
\usepackage{comment}
\usepackage[export]{adjustbox}
\usepackage{arydshln}
\usepackage{multirow, multicol}
\usepackage[ruled, vlined, linesnumbered]{algorithm2e}

\SectionNumbersOn    

\usepackage[font=footnotesize,labelfont=bf,labelsep=period,width=0.95\textwidth]{caption}   
\usepackage[font=scriptsize,labelfont=bf]{subcaption}                     
\usepackage[capitalize]{cleveref}
\crefname{figure}{Figure}{Figures}
\Crefname{figure}{Figure}{Figures}
\crefname{table}{Tab.}{Tabs.}
\Crefname{table}{Table}{Tables}
\crefname{equation}{Eq.}{Eqs.}
\Crefname{equation}{Eq.}{Eqs.}
\crefname{section}{Sec.}{Secs.}
\Crefname{section}{Section}{Sections}
\usepackage{braket}

\usepackage{etoolbox}

\makeatletter
\renewcommand*{\acs@author@fnsymbol@symbol}[1]{
    \ifcase #1 *\or
    1\or
    {\#}\or
    2\or
    3\or
    4\or
    5\or
    6\or
    7\or
    8\or
    9
    \fi
}
        
\renewcommand*\acs@contact@details{
    {\sffamily *\,Corresponding author; E-mail: \acs@email@list }%
    \acs@number@list
}           

\patchcmd{\acs@address@list@auxii}
{\acs@author@fnsymbol{\acs@affil@marker@cnt}}
{\textsuperscript{\acs@author@fnsymbol{\acs@affil@marker@cnt}}}
{}{}

\patchcmd{\acs@address@list@auxii}
{{\acs@author@fnsymbol{\acs@affil@marker@cnt}\@nameuse{@altaffil@\@roman\@tempcnta}\par}}
{{\textsuperscript{\acs@author@fnsymbol{\acs@affil@marker@cnt}}\@nameuse{@altaffil@\@roman\@tempcnta}\par}}
{}{}        

\makeatother    

\title[CCvsDMRGvsCI]{Definitive Assessment of the Accuracy, Variationality, and Convergence of Relativistic Coupled Cluster and Density Matrix Renormalization Group in 100-Orbital Space}

\author{Shiv Upadhyay}
\affiliation[University of Washington]
{Department of Chemistry, University of Washington, Seattle, WA 98195, USA}
\alsoaffiliation[co-first]
{Authors contributed equally to this work}

\author{Agam Shayit}
\affiliation[Physics]
{Department of Physics, University of Washington, Seattle, WA 98195, USA}
\alsoaffiliation[co-first]
{Authors contributed equally to this work}

\author{Tianyuan Zhang}
\affiliation[University of Washington]
{Department of Chemistry, University of Washington, Seattle, WA 98195, USA}
\alsoaffiliation[co-first]
{Authors contributed equally to this work}

\author{Stephen H. Yuwono}
\affiliation{
             Department of Chemistry and Biochemistry,
             Florida State University,
             Tallahassee, FL 32306-4390, USA}
             
\author{A. Eugene DePrince III}
\affiliation{
             Department of Chemistry and Biochemistry,
             Florida State University,
             Tallahassee, FL 32306-4390, USA}

\author{Xiaosong Li}
\affiliation[University of Washington]
{Department of Chemistry, University of Washington, Seattle, WA 98195, USA}
\alsoaffiliation[Physics]
{Department of Physics, University of Washington, Seattle, WA 98195, USA}
\email{xsli@uw.edu}
\begin{document}

\def\thepage{\arabic{page}}
\def\thetable{\arabic{table}}
\def\thefigure{\arabic{figure}}
\def\thesection{\arabic{section}}
\def\theequation{\arabic{equation}}

\pagebreak

\begin{abstract}
Accuracy, variationality, and convergence underpin the reliability of modern electronic structure methods, yet \emph{definitive} benchmarks in the relativistic regime remain elusive due to the absence of numerically exact full configuration interaction (CI) references. Recent algorithmic advances in the CI framework, enabled by the small-tensor-product (STP) decomposition approach, have dramatically extended the tractable size of the configuration space, making numerically exact CI calculations feasible in large active spaces previously beyond reach. In this work, we employ the recently developed STP-CI framework to perform large-scale numerically exact CI calculations and directly benchmark relativistic coupled cluster and density matrix renormalization group methods. \emph{Definitive} benchmarking of approximate relativistic electronic structure methods is ensured through the application of the gap theorem, which provides rigorous error bounds on the CI reference and establishes a controlled standard for assessing accuracy, variationality, and convergence.
\end{abstract}

\section{Introduction}

Accuracy, variationality, and convergence are foundational criteria that determine the reliability and predictive power of modern electronic structure methods in quantum chemistry. While full configuration interaction (FCI) provides the formally exact description of the electronic structure of a chemical system within a given basis set,\cite{Shavitt77_book,Shavitt98_3,Schaefer99_143,Carsky02_book,Shavitt03_book,Shepard12_108} 
it is generally intractable for all but the smallest chemical systems. 

Many alternative wavefunction-based methods have been developed to overcome the prohibitive size of FCI. These approaches seek approximate solutions to the FCI problem by introducing controlled truncations of the parameter space based on different types of electron correlation.

Correlation energy for molecules is often conceptually partitioned into two components \cite{Musial07_291}:
\begin{enumerate}
\item \emph{Dynamic correlation}, which accounts for the short-range electron--electron repulsion beyond the mean-field and arises from many small contributions spread over a large number of determinants;
\item \emph{Static (or nondynamic) correlation}, which originates from near-degenerate electronic configurations and is dominated by large contributions from a relatively small set of determinants.
\end{enumerate}

Two widely used classes of methods for approximating FCI, based on different treatments of electron correlation, are coupled cluster (CC) and density matrix renormalization group (DMRG). Both methods are post-Hartree--Fock methods, generating an approximated wavefunction solution from a mean-field reference Slater determinant. We refer the readers to  Refs.~\citenum{Musial07_291,Bartlett12_126} for a thorough review of the CC family of methods and Refs.~\citenum{sharma11_465,reiher20_040903} for DMRG.

In CC methods, dynamic correlation is recovered on top of the mean-field reference by including excited configurations in the form of an {\it exponential ansatz}:\cite{Cizek66_4256}
\begin{equation}\label{eq:ccExpAnsatz}
    \ket{\Psi} = e^{\hat{T}}\ket{\Phi} = e^{\hat{T}_1+\hat{T}_2+\cdots}\ket{\Phi},
\end{equation}
where $\Phi$ is the Hartree--Fock determinant and $T_n$ is the $n$-body cluster operator, defined as
\begin{equation}\label{eq:ccClusterOps}
    \hat{T}_n\equiv \left(\frac{1}{n!}\right)^2 \sum^n_{ij\cdots ab\cdots} t_{ij\cdots}^{ab\cdots} \hat{a}^\dagger \hat{b}^\dagger \cdots \hat{j}\hat{i},
\end{equation}
where $t_{ij\cdots}^{ab\cdots}$ are cluster amplitudes and $\hat{a}^\dagger \hat{b}^\dagger \cdots$ and $\hat{i}\hat{j} \cdots$ are creation and annihilation operators on unoccupied $ab\cdots$ and occupied $ij\cdots$ orbitals, respectively. 

Often, the expansion is truncated at a particular excitation level, with the most common choices being $\hat{T}=\hat{T_1}+\hat{T_2}$ and $\hat{T}=\hat{T_1}+\hat{T_2}+\hat{T_3}$, resulting in the {\color{black} CC with} singles and doubles (CCSD)\cite{Bartlett82_1910,Zerner82_4088} and {\color{black} CC with} singles, doubles, and triples (CCSDT)\cite{Bartlett87_7041,Schaefer88_382} methods.
Methods of greater accuracy than CCSD and smaller computational footprint than CCSDT have been developed, among which {\color{black}are} the CCSD(T) method{\color{black}
\cite{Head-Gordon89_479}
and the completely renormalized (CR) CC(2,3) approach.\cite{Wloch05_224105,Gour06_2149}}
In the CCSD(T) method, after a converged CCSD calculation, the correlation energy is improved by a one-shot 
triples-correction without explicit storage of the $t_{ijk}^{abc}$ amplitudes\cite{Head-Gordon89_479,Noga90_513}.
We refer readers to Ref.~\citenum{Head-Gordon89_479} for details in this non-iterative approximation.
{\color{black} In the CR-CC(2,3) approach, one constructs the triples energy corrections to CCSD in a non-perturbative manner via the method-of-moments of CC equations,\cite{Kowalski00_1,Piecuch00_18,Piecuch00_5644} which involves solving the left-hand (``lambda'') CCSD equations as well. The resulting implementation of non-iterative CR-CC(2,3) corrections is similar in overall structure with that of CCSD(T). Furthermore, it is well known that CR-CC(2,3) tends to outperform CCSD(T) in recovering correlation energies, while remaining numerically stable in situations where static correlations become more important (\emph{e.g.}, in bond breaking).\cite{Wloch05_224105,Gour06_2149,DePrince24_6521}}

All CC methods mentioned above are in the category of single-reference methods, in which dynamic correlation is described well by the exponential ansatz. However, their single-reference nature is not efficient in describing static correlation.
Many multi-reference CC methods {\color{black} have} been reported to describe static and dynamic correlation equally well\cite{ Bartlett12_182, Paldus17_477, Evangelista18_030901},
{\color{black} and there exist unconventional single-reference CC approaches that are designed to handle such situations (see, e.g., Refs.~\citenum{Paldus17_477,Piecuch22_e2057365} and the references therein),} but they are beyond the scope of this work.

Another common alternative to FCI is the DMRG method, in which the wavefunction is represented by a tensor network\cite{reiher20_040903, sharma11_465, vanneck14_272}.
Tensor-network wavefunctions are factorizations of the exponentially-large FCI coefficient tensor that come in various flavors, including the matrix product state (MPS)\cite{reiher20_040903, sharma11_465, vanneck14_272}, the tree tensor network (TTN) state\cite{vidal09_165129, noack10_205105, chan13_134113, vidal06_022320, vidal09_235127}, and the projected entangled pair states (PEPS)\cite{cirac08_143}. The DMRG algorithm allows for the efficient variational optimization of MPS wavefunctions. The MPS ansatz is given by 
\begin{equation}\label{eq:MPS}
\footnotesize
    \ket{\Psi} = \sum_{\sigma}^{N}\sum_{a_1}^{m}\cdots\sum_{a_{N-1}}^{m} M^{\sigma_1}_{1,a_1} M^{\sigma_2}_{a_1,a_2} \cdots M^{\sigma_N}_{a_{N-1},1} \ket{\sigma_1,\sigma_2,\cdots,\sigma_N},
\end{equation}
where $\sigma$ are sites with $\ket{\sigma_1,\sigma_2,\cdots,\sigma_N}$ representing an occupation vector, $N$ is the number of sites, $M$ are site tensors with auxiliary indices $a$ that are truncated at a given bond dimension $m$. 

In quantum chemical applications, molecular orbitals can be mapped directly onto sites, making DMRG a particularly attractive framework for electronic structure problems, especially for the accurate treatment of static correlation. Moreover, DMRG exhibits polynomial scaling with a compact variational ansatz and systematically controllable convergence toward the FCI limit through increases in the bond dimension $m$.

The practical validity of approximate electronic structure methods such as CC and DMRG is often assessed through comparisons with experimental observables.\cite{dixon01_3484, barone10_637,sherrill11_194102,sundholm11_2473,dixon12_2381,szalay14_3757,  Li14_164116, Li15_4146,pierloot16_4352,Li19_6617,izsak20_564,neese20_90, martin21_8987,Li22_5011, martin22_25555, Li24_3408,Li24_6521,Li25_084110,Li25_104112,ma25_164115,Li26_016101} Such benchmarks, however, cannot disentangle errors arising from methodological approximations from those due to the finite atomic basis sets employed. A more controlled assessment is therefore obtained by benchmarking computed observables against FCI results, which provide a well-defined theoretical reference within a given basis.

Despite its importance, systematic and controlled benchmarking of approximate electronic structure methods in the relativistic regime remains largely unexplored, owing to the rapidly expanding configuration space associated with complex-valued two- and four-component wavefunctions and dense spinor manifolds. 
The recent methodological advances in numerically exact FCI methods based on the STP decomposition\cite{Li24_041404,Li25_11016}  have substantially extended the tractable orbital space, enabling controlled, deterministic benchmarks at system sizes that were previously inaccessible. These developments now make it possible to establish definitive reference data for assessing the accuracy, variationality, and convergence of approximate relativistic electronic structure methods in large active spaces. 

In this work, we present definitive large-scale benchmarks of relativistic CC and DMRG against numerically exact CI references in large active spaces. We consider three benchmark systems that exhibit pronounced relativistic effects and a challenging interplay of dynamic and static electron correlation. To rigorously quantify the reliability of the CI reference, we employ the gap theorem\cite{DavisKahan70_1,Parlett98_book,Knyazev13_244} to establish a lower bound on the CI energy, ensuring a definitive assessment of the approximate relativistic electronic structure methods. 

\section{Theory}

A key advantage of using numerically exact CI as the reference is the ability to compute the Ritz residual of the CI eigenvector exactly, which in turn enables rigorous error bounds on the true Hamiltonian eigenvalue within the given basis\cite{Minkoff05_90}. Of these bounds, the gap theorem\cite{DavisKahan70_1,Parlett98_book,Knyazev13_244}  offers the tightest lower bound, which is computed using the Ritz eigenvector and eigenvalue as
\begin{equation}
|\delta E| \leq \frac{\|\mathbf{r}\|^2}{\gamma_0},
\end{equation}
where $\mathbf{r}$ is the Ritz residual of the desired state and the gap $\gamma_0 \equiv E_1-\tilde{E}_0$ is the difference between $E_1$, the (unknown) exact energy of the first excited state, and $\tilde{E}_0$, the computed Ritz value of the ground state. 
Because $\gamma_0$ is unknown, one can estimate its order of magnitude using approximate methods or use experimental values to compute a surrogate for the true gap. This is often sufficient for well-separated, for which $\gamma_0$ is much larger than reasonable values $\|\mathbf{r}\|^2$ attains after a few Davidson iterations\cite{Davidson75_87}.

In more pathological cases, such as ground-state near-degeneracy, one can instead obtain an exact lower bound on the gap $\gamma_0$ by including the posterior error bound of the first excited state~\cite{yosida1995_book,saad2011_book,Knyazev13_244} in the Davidson calculation:\cite{Minkoff05_90}
\begin{equation}\label{eq:gammaLowerBound}
    \gamma_0 = E_1-\tilde{E_0}\geq \left(\tilde{E_1}-\|\mathbf{r_1}\|\right)-\tilde{E}_0 \equiv \gamma^-_0,
\end{equation}
where $\mathbf{r_1}$ is the residual associated with the Ritz value $\tilde{E}_1$. Combined with the variational bound associated with all Krylov subspace methods\cite{Minkoff05_90},
\begin{equation}\label{eq:upperBound}
    E_0 \leq \tilde{E}_0,
\end{equation}
one can place tight upper and lower bounds on the true eigenvalue:
\begin{equation}\label{eq:upperLowerBounds}
    \tilde{E}_0 - \frac{\|\mathbf{r}\|^2}{\gamma_0}\leq E_0 \leq \tilde{E}_0.
\end{equation}


The ability to obtain rigorous error bounds for large-scale CI calculations enables precise benchmarking of electronic structure methods at scales that were previously intractable, extending well beyond small numbers of correlated orbitals and electrons. These bounds explicitly bracket the true energy eigenvalue, thereby allowing approximate methods to be evaluated faithfully across different correlation regimes.

\section{Results and Discussion}

The benchmark set consists of three systems of varying degrees of symmetry and correlation: HBrTe, Rb$_4$, and Xe$_2$:
\begin{itemize}
    \item HBrTe (100 two-spinor orbitals, 88 electrons, x2c-TZVPall\cite{Weigend17_3696}) is a substituted form of a hydrogen chalcogenide where one of the hydrogens was substituted with a bromine atom to decrease the symmetry to the molecule.
    \item Square Rb$_4$ (50 two-spinor orbitals, 28 electrons, cc-pVTZ-x2c\cite{Peterson17_244106}), a relativistic analogue of H$_4$\cite{Stevens73_3378,Clary96_8413,Hellgren24_074106,Noe21_124108,Nakatsuji23_140359,Champagne12_024315,Noga11_418,Fromager22_032203,Ellis11_124108,Head-Gordon00_8873,Sorella19_084102,Whaley23_030307,Mazziotti13_44}, displays strong static correlation.
    \item Xe$_2$ (60 two-spinor orbitals, 12 electrons, x2c-TZVPall-2c\cite{Weigend17_3696}) is a dynamically correlated noble gas dimer\cite{Saue12_54,Hobza96_425,Yang97_7921,Evangelisti03_303,Malijevsky03_2102,Perdew05_114102,Stoll05_3917,Becke09_719,Angyan10_244108}.
\end{itemize}

{\color{black} The details of the numerically exact relativistic configuration interaction calculations are given in Ref.~\citenum{Li25_11016}, and the reference values generated in that work are used here. }
Relativistic coupled-cluster (CC) calculations were performed using the Chronus Quantum software package.\cite{Li19_6617, Li21_5438, Li23_044113, Li24_3408, Li24_6521, Li25_084110, Li25_104112, Li26_016101, Li20_e1436} Relativistic density matrix renormalization group (DMRG) calculations were carried out with the X2C-DMRG implementation in Block2,\cite{Li22_5011, Li24_6521} using one- and two-electron integrals generated by Chronus Quantum. All configuration interaction (CI) calculations employed the STP distributed active space (STP-DAS)\cite{Li24_041404,Li25_11016} algorithm as implemented in Chronus Quantum.

Relativistic effects were incorporated variationally within the Dirac formalism using the one-electron exact two-component (1eX2C) method along with a Dirac--Coulomb--Breit-parameterized
Boettger factor\cite{Li23_5785}. The 1eX2C method uses the one-electron portion of the Hamiltonian to approximately reduce the four-component Dirac equation into a two-component eigenvalue problem for the positive energy (electronic) sector of the Hamiltonian\cite{Dyall97_9618,Dyall98_4201,Dyall99_10000,Dyall01_9136,Liu05_241102,Peng06_044102,Cheng07_104106,Saue07_064102,Peng09_031104,Liu10_1679,Liu12_154114,Reiher13_184105,Li16_104107,Li16_3711,Repisky16_5823,Li17_2591,Cheng21_1536,Li22_2947, Li22_2983,Li22_5011}.

All computations were performed on the National Energy Research Scientific Computing Center (NERSC) Perlmutter supercomputer (AMD EPYC 7763 Milan processors, 128 cores per node, 512 GB memory per node, and 200 GB$\cdot$s$^{-1}$ interconnect).  The benchmark set consists of ground-state energy calculations for HBrTe, Rb$_4$, and Xe$_2$.  The largest X2C-DMRG calculation (HBrTe with 100 two-spinor orbitals and 88 electrons at bond dimension $m=1000$) required 30 nodes, whereas the corresponding X2C-CC calculation was performed on 4 nodes. The largest X2C-CI calculation ($10^{15}$ complex-valued determinants) was carried out on a total of 1,000 nodes.{\color{black}\cite{Li25_11016}}

\begin{table}[H]
    \footnotesize
    \captionsetup{width=\columnwidth}
    \caption{The discrepancy (in Hartree), $\Delta E=E_\text{CC}-E_\text{CI}$, between X2C-CC ground state energies and the corresponding X2C-CI values for HBrTe (100 2-spinor orbitals, 88 electrons, x2c-TZVPall\cite{Weigend17_3696}), Rb$_4$ (50 2-spinor orbitals, 28 electrons, cc-pVTZ-x2c\cite{Peterson17_244106}), and Xe$_2$ (60 2-spinor orbitals, 12 electrons, x2c-TZVPall-2c\cite{Weigend17_3696}).}
    \label{tab:CCBenchmarks}
    \begin{tabular}{l|rrr}
    & \multicolumn{1}{c}{HBrTe} & \multicolumn{1}{c}{Rb$_4$} & \multicolumn{1}{c}{Xe$_2$}\\\cmidrule(r){1-4}
    X2C-CCSD & $1.26\times10^{-3}$ & $7.06\times10^{-4}$ & $2.52\times10^{-3}$ \\
    X2C-CCSD(T) & $1.97\times10^{-4}$ & $2.27\times10^{-4}$ & $3.41\times10^{-4}$\\
    \color{black}X2C-CR-CC(2,3) & \color{black}$-1.25\times10^{-4}$ & \color{black}$1.68\times10^{-4}$ & \color{black}$-1.19\times10^{-4}$\\
    X2C-CCSDT & $1.62\times10^{-5}$ & $1.61\times10^{-4}$ & $-4.73\times10^{-5}$\\
    \bottomrule
    \end{tabular}
\end{table}

\begin{figure}
    \centering
    \captionsetup{width=\linewidth}\includegraphics[width=0.8\linewidth]{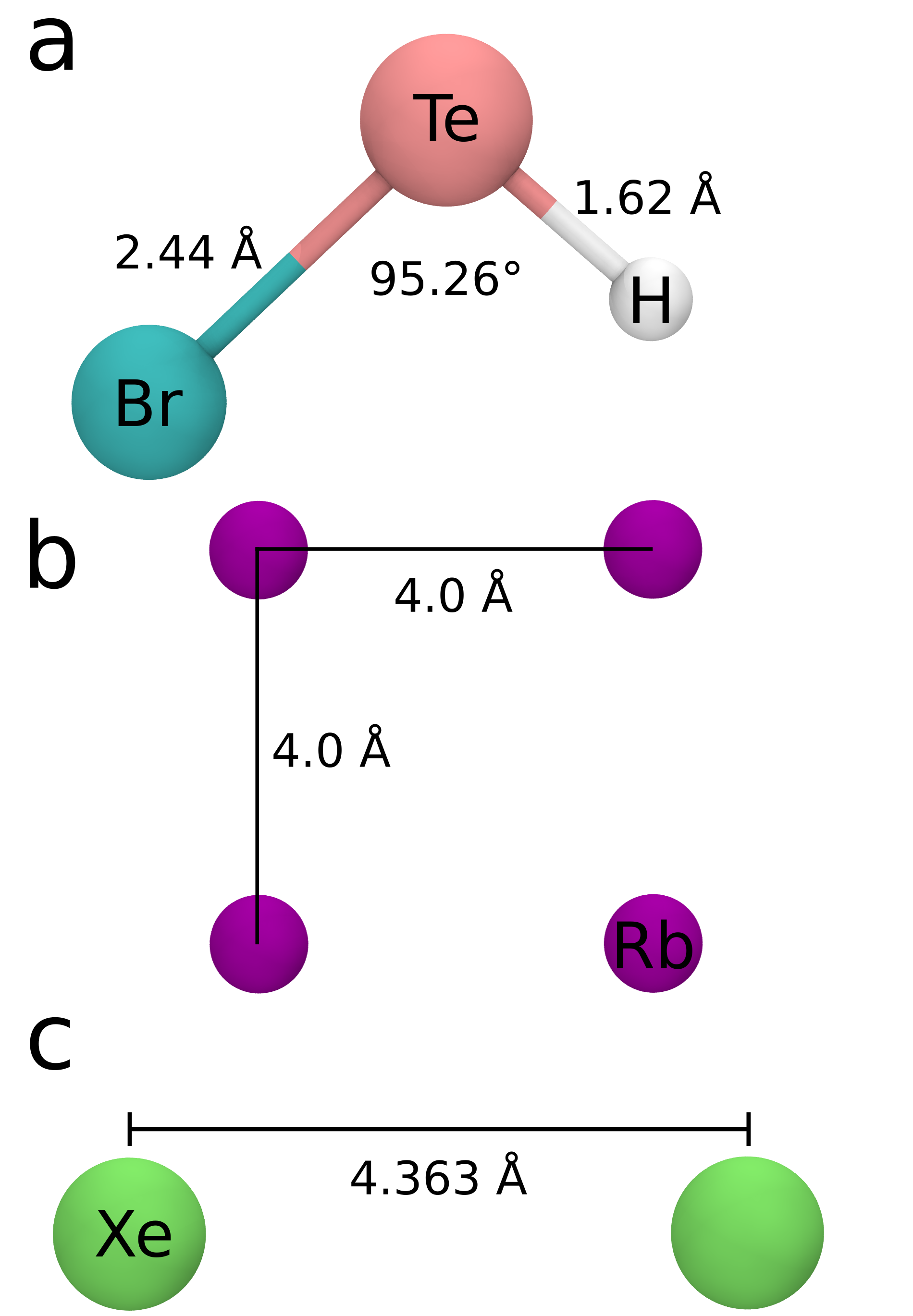}
    \caption{Benchmark chemical systems used in this work: (a) HBrTe an asymmetric modification of H$_2$Te (b) Rb$_4$ a relativistic analogue of the H$_4$ square planar system which displays strong static correlation (c) Xe$_2$ a dispersion bound noble gas complex dominated by dynamic correlation. }
    \label{fig:systems}
\end{figure}

\Cref{tab:CCBenchmarks} compares X2C-CC {\color{black}ground state} energies against the corresponding X2C-CI reference values. All CC levels achieve chemical accuracy (1.58 millihartree, 1 kcal/mol) relative to the CI limit; however, none reach microhartree-level precision. Such higher accuracy is essential when resolving individual wavenumbers, as required for high-resolution spectroscopic simulations.

Unsurprisingly, the most accurate energies are obtained by X2C-CCSDT, which incorporates triple-excitation amplitudes converged iteratively rather than one-step perturbatively (as is the case with X2C-CCSD(T)). {\color{black} Furthermore, the CR-CC(2,3) triples correction outperforms CCSD(T) in reproducing the parent X2C-CCSDT energetics, although it overcorrelates two of the test systems with respect to the X2C-CI limit by about 0.1 millihartree.} Notably, the error in the X2C-CCSDT ground state energy of Rb$_4$ is an order of magnitude larger than for the other two systems. This behavior is expected because {\color{black} conventional CC} is inherently a single-reference method. Because the ground state of Rb$_4$ is a relativistic analogue of the H$_4$ system\cite{Jordan17_074106}, it is highly multireference, and {\color{black} CC} will not accurately capture its static correlation.

Additionally, these benchmarks showcase the non-variational nature of {\color{black} CC}-based methods. This is evident in the negative entry in \Cref{tab:CCBenchmarks} corresponding to the X2C-CCSDT ground state energy of Xe$_2$. 
Applying the energy-bound analysis (\Cref{eq:upperLowerBounds}) shows that the computed X2C-CI energy is bound from below by the true eigenvalue of the Hamiltonian matrix\cite {Li25_11016}:
\begin{equation}\label{eq:xe2Bounds}
    0 \leq \tilde{E}_0 - E_0\leq 1.76\times 10^{-5}.
\end{equation}
This analysis also indicates that the X2C-CCSDT energy lies below the true eigenvalue of the Hamiltonian matrix by at least 30~$\mu E_\text{h}$, nicely demonstrating the nonvariational nature of {\color{black} CC}-based methods.

\begin{table}[H]
    \footnotesize
    \captionsetup{width=\linewidth}
    \caption{The discrepancy (in Hartree), $\Delta E=E_\text{DMRG}-E_\text{CI}$,  between X2C-DMRG ground state energies and the corresponding X2C-CI values at various bond dimensions. The systems are HBrTe (100 2-spinor orbitals, 88 electrons, x2c-TZVPall\cite{Weigend17_3696}), Rb$_4$ (50 2-spinor orbitals, 28 electrons, cc-pVTZ-x2c\cite{Peterson17_244106}), and Xe$_2$ (60 2-spinor orbitals, 12 electrons, x2c-TZVPall-2c\cite{Weigend17_3696}).}
    \label{tab:DMRGBenchmarks}
\begin{tabular}{c|ccccc}
$m$ & HBrTe & Rb$_4$ & Xe$_2$ \\\cmidrule(r){1-6}

$m=50$ & $4.01\times10^{-3}$ & $1.44\times10^{-4}$ & $3.82\times10^{-2}$ \\

$m=100$ & $1.72\times10^{-3}$ & $4.99\times10^{-5}$ & $1.22\times10^{-2}$ \\

$m=200$ & $6.31\times10^{-4}$ & $1.63\times10^{-5}$ & $9.95\times10^{-3}$ \\

$m=500$ & $1.50\times10^{-4}$ & $2.08\times10^{-6}$ & $5.58\times10^{-3}$ \\

$m=1000$ & $5.24\times10^{-5}$ & $5.39\times10^{-7}$ & $2.40\times10^{-3}$ \\

\bottomrule
\end{tabular}
\end{table}

\Cref{tab:DMRGBenchmarks} compares the computed X2C-DMRG energies with the corresponding X2C-CI benchmark values. The results clearly demonstrate the variational nature of DMRG. As the bond dimension increases, the energies decrease monotonically and converge toward the benchmark. In the limit of infinite bond dimension, DMRG recovers the exact solution of the Hamiltonian eigenvalue problem. {\color{black} In practice, it is not always feasible to reach the infinite bond dimension limit as the results for Xe$_2$ demonstrate. }

In contrast to the {\color{black} CC} results in \Cref{tab:CCBenchmarks}, DMRG performs substantially better for Rb$_4$, a multireference system dominated by static correlation, than for Xe$_2$, which is primarily governed by dynamic correlation. This behavior arises because a finite bond dimension constrains orbital entanglement, limiting DMRG's ability to capture the weak, long-range correlations characteristic of dynamic correlation.

\begin{figure*}
    \centering
    \captionsetup{width=\linewidth}\includegraphics[width=\linewidth]{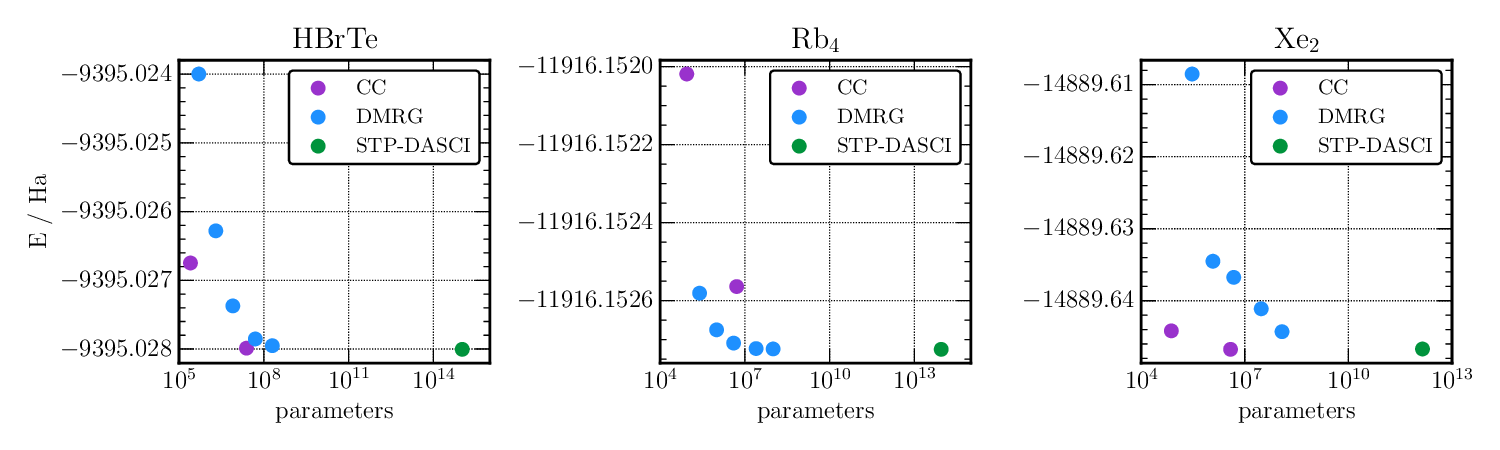}
    \caption{The number of parameters in the various wave function ansatzes used in this work: DMRG, CC, and STP-DASCI for the three benchmark systems. }
    \label{fig:nnzcoeffs}
\end{figure*}
\begin{figure}
    \centering
    {\tiny HBrTe} \\
    \captionsetup{width=\linewidth}\includegraphics[width=\linewidth]{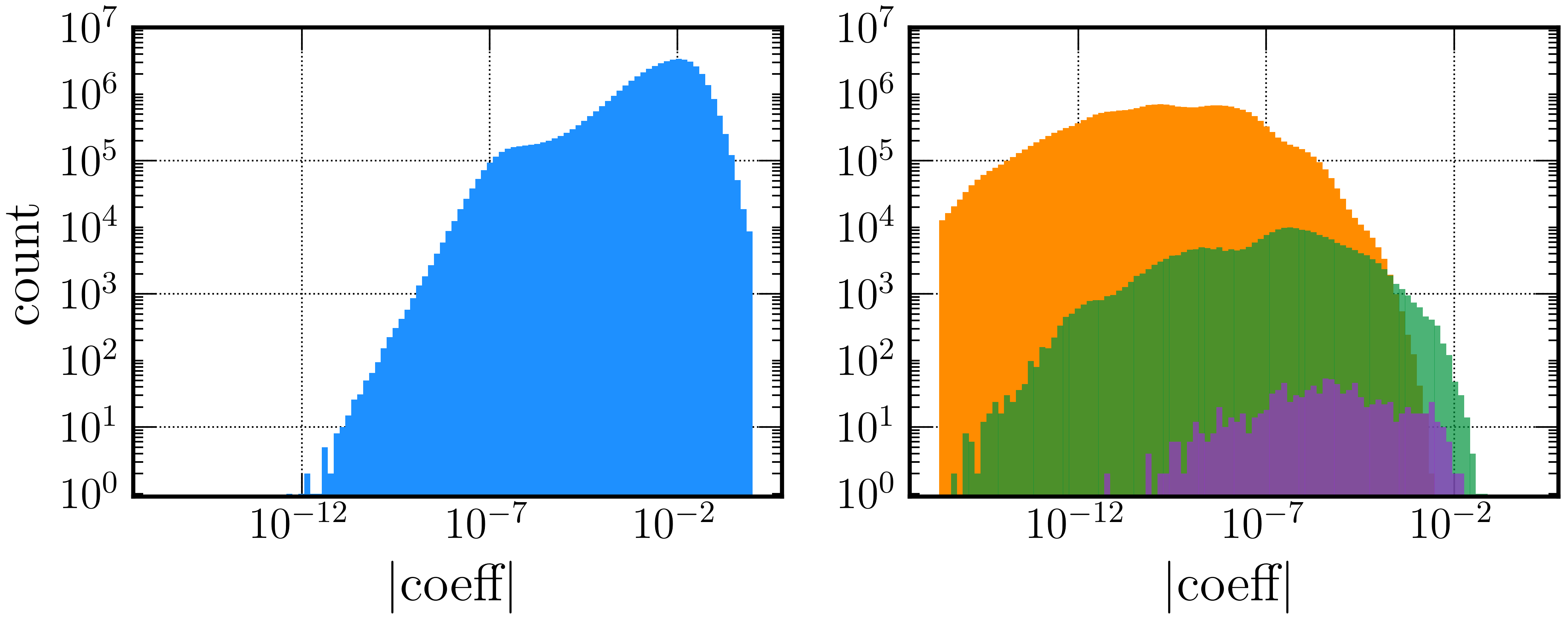}
    {\tiny Rb$_4$} \\
    \captionsetup{width=\linewidth}\includegraphics[width=\linewidth]{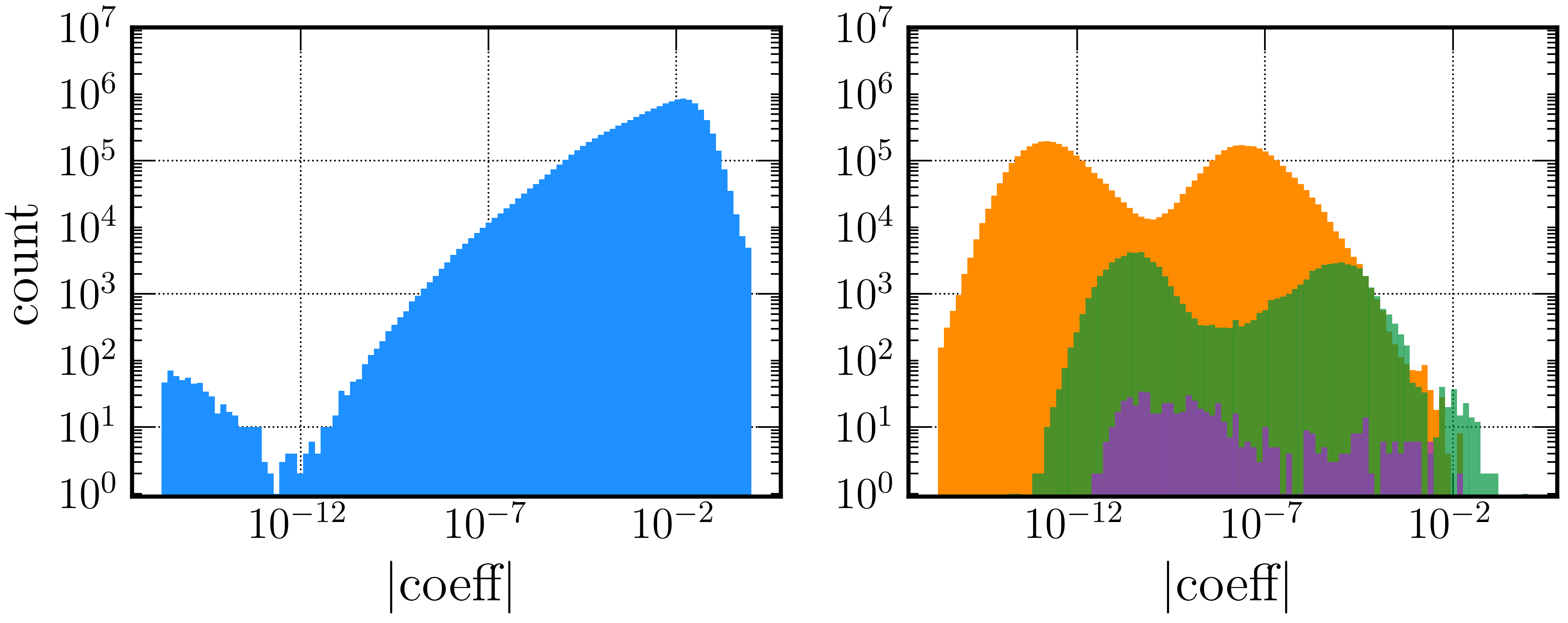}
    {\tiny Xe$_2$} \\
    \captionsetup{width=\linewidth}\includegraphics[width=\linewidth]{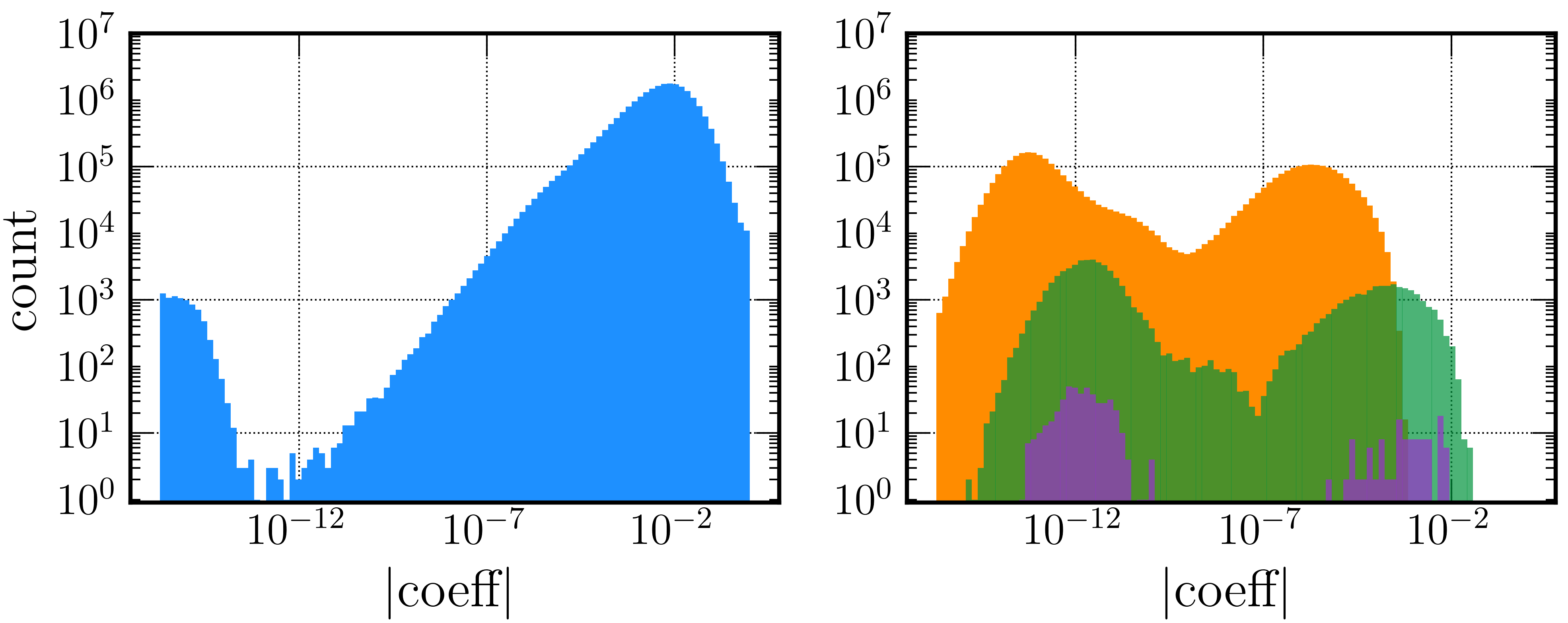}
    \caption{A histogram of the absolute value of the coefficients in the (left blue) DMRG matrix product state ansatz and the (right) CCSDT exponential ansatz (where purple, green and orange represent amplitudes in $\hat{T}_1$, $\hat{T}_2$ and $\hat{T}_3$, respectively). }
    \label{fig:hist}
\end{figure}

{\color{black}
In order to better understand the behavior of the various wavefunction ansatzes, we study the number of elements of the various wavefunction parameterizations: CI coefficients for the numerically exact CI calculations, MPS parameters for the DMRG calculations, and cluster amplitudes for the CC calculations.
\Cref{fig:nnzcoeffs} plots the convergence of the energy with respect to the number of parameters.
As was observed in the energetic convergence, the DMRG calculations converge smoothly for HBrTe and the statically correlated Rb$_4$.
For the system that DMRG struggled with, the dynamically correlated Xe$_2$ dimer, the convergence with increase parameter count is erratic demonstrating that the MPS wavefunction struggles to capture the many important low weight
configurations characteristic of dynamically correlated systems.
The CC calculations and DMRG calculations display a similar convergence for HBrTe, but CC struggles with the statically correlated Rb$_4$ rather than the dynamically correlated Xe$_2$. 
This is manifested by the convergence of the CC calculations display nearly opposite trends to the DMRG calculations on these systems.

To assess the strengths and limitations of the wavefunction ansatz compactness, we analyze the distribution of its parameters by constructing a histogram, as shown in \cref{fig:hist}.
Since the matrix product state wavefunction and the exponential {\color{black} CC} wavefunction are fundamentally different in construction, the direct comparison of these distributions is not relevant. However the relative distributions between systems is physically meaningful. 
For example, the CCSDT amplitudes, which contains singles, doubles and triples amplitudes, for HBrTe are present as a single distribution with multiple distinct peaks.
The distributions of amplitudes show bimodal patterns for Rb$_4$ and Xe$_2$.
This is because the bond lengths in these systems are much longer than in the HBrTe system.
Consequently, cluster amplitudes are divided into intra- and inter-atomic amplitudes, with the latter orders of magnitude smaller than the former.

\begin{table}[H]
    \footnotesize
    \captionsetup{width=\columnwidth}
    \caption{The $T_1$, $D_1$, $D_2$ diagnostic\cite{Schaefer89_81,Taylor89_199,Nielsen98_423,Janssen99_568,Laestadius23_9106} and the maximum $\hat{T}_2$ amplitude ($\max~t_2$) diagnostic for HBrTe (100 2-spinor orbitals, 88 electrons, x2c-TZVPall\cite{Weigend17_3696}), Rb$_4$ (50 2-spinor orbitals, 28 electrons, cc-pVTZ-x2c\cite{Peterson17_244106}), and Xe$_2$ (60 2-spinor orbitals, 12 electrons, x2c-TZVPall-2c\cite{Weigend17_3696}) from corresponding CCSD cluster amplitudes.}
    \label{tab:CCdiagnostic}
    \begin{tabular}{c|ccr}
    Diagnostic & \multicolumn{1}{c}{HBrTe} & \multicolumn{1}{c}{Rb$_4$} & \multicolumn{1}{c}{Xe$_2$}\\\cmidrule(r){1-4}
    $T_1$ & 0.0048 & 0.0049 & 0.0109 \\
    $D_1$ & 0.0375 & 0.0179 & 0.0148 \\
    $D_2$ & 0.1076 & 0.7691 & 0.0656 \\
    $\max~t_2$ & 0.0606 & 0.7233 & 0.0294 \\
    \bottomrule
    \end{tabular}
\end{table}

Indicators for the reliability of single-reference CC calculation, including the $T_1$, $D_1$, $D_2$ diagnostics\cite{Schaefer89_81,Taylor89_199,Nielsen98_423,Janssen99_568,Laestadius23_9106} and the maximum $\hat{T}_2$ amplitude ($\max~t_2$) diagnostic, are listed in \cref{tab:CCdiagnostic}.
As above discussion has shown, Rb$_4$ is the most challenging system for CC theory among the three systems investigated in this paper, we expect diagnostic values of Rb$_4$ to be the largest among three molecules.
However, only diagnostics based on double substitution ($D_2$ and $\max~t_2$) indicate multi-reference nature of the Rb$_4$ system.
We are also able to observe the presence of a single large $\hat{T}_2$ amplitude in the middle right panel of \cref{fig:hist}.
Rb$_4$ is one more case where single-substitution-based diagnostics ($T_1$ and $D_1$) fail.
Nevertheless, Rb$_4$ is analogous to H$_4$ with core electrons and, thus, can be expected to be well-described if quadruple excitation is included in the CC calculation.
Moving on, in each of the right panels in \cref{fig:hist}, the large-value tail is more pronounced in $\hat{T}_2$ than in $\hat{T}_1$, indicating significant electronic correlations (consistent with the usual many-body perturbation analysis of these cluster components). There are also relatively large-valued $\hat{T}_3$ amplitudes ($\approx10^{-4}$ to $10^{-2}$) that are comparable in magnitude to parts of $\hat{T}_1$ or $\hat{T}_2$, especially for Rb$_4$, which indicates non-negligible static correlation effects in the systems we examine here (cf.~how different ranks of cluster operator behave in the strong correlation regime of a model Hamiltonian in Ref.~\citenum{Scuseria16_125124}). The Xe dimer shows a sharp drop in the large-value part of the $\hat{T}_3$ distribution compared to $\hat{T}_1$ or $\hat{T}_2$, again reinforcing our observation that correlation is mostly dynamical in this system.

The MPS coefficients from the ($m$=1000) DMRG calculation shed light into why DMRG struggles with the dynamic correlation in this system. 
For all three systems, the MPS coefficients vanish beyond roughly 1$\times$10$^{-8}$.
This can be understood as a truncation of the optimized tensors during the DMRG sweeps based on a threshold of the singular values of the tensors.
This works well for systems such as HBrTe and Rb$_4$ where these low weight contributions are not relevant.
For heavy-tailed wavefunctions such as the dynamically correlated Xe dimer, the many low weight excitations characteristic of dynamic correlation are lost to truncation.
Finally, it is important to note that the lack of dynamic correlation in these DMRG calculations is not caused by a small active space, which is often discussed in the literature and addressed by methods such as perturbative treatments on top of a DMRG wavefunction.
In this work, DMRG is neglecting dynamic correlation within the active space, which can only be addressed by going to larger bond dimensions or possibly by other techniques such as augmenting smaller bond dimension DMRG calculations with machine learning.\cite{Veis15_3295}
}

\section{Conclusions}

The ability to perform extremely large-scale, numerically exact CI calculations, enabled by categorical compression within the STP-DAS CI framework, now permits precise benchmarking of the accuracy and convergence behavior of wavefunction-based electronic structure methods at scales that were previously intractable. In this work, we carry out such benchmarks for two widely used relativistic approaches: coupled cluster (X2C-CC) and density matrix renormalization group (X2C-DMRG). The benchmark systems (HBrTe, Rb$_4$, and Xe$_2$) span a range of relativistic effects, including spin--orbit coupling, point-group symmetry, and varying degrees of dynamic and static electron correlation.

The {\color{black} CC} benchmarks demonstrate its well-known strength in describing dynamic correlation in single-reference systems. At the highest level considered, X2C-CCSDT reproduces the CI reference energies of the single-reference systems HBrTe and Xe$_2$ to within a few wavenumbers. In contrast, performance degrades by more than an order of magnitude for the multireference system Rb$_4$, consistent with the intrinsic limitations of single-reference CC theory. Moreover, the non-variational nature of {\color{black} CC} is conclusively established by showing that the X2C-CCSDT energy of Xe$_2$ lies at least $30~\mu E_\mathrm{h}$ below the true Hamiltonian eigenvalue within the given basis.
{\color{black} From a purely electronic correlation consideration, it would be interesting to apply single-reference CC methodologies tailored for strong correlations (see, \emph{e.g.}, Refs.~\citenum{Paldus17_477, Piecuch22_e2057365} and the references therein) within the relativistic framework and examine their systematic convergence toward exact CI reference data.}

The DMRG benchmarks exhibit the expected convergence behavior characteristic of tensor-network methods. DMRG efficiently recovers the CI energy of Rb$_4$, which is dominated by static correlation, even at modest bond dimensions. By contrast, for dynamically correlated systems such as HBrTe and Xe$_2$, convergence toward the CI limit remains slow, and even large bond dimensions fail to achieve the level of accuracy typically required in high-precision quantum chemical applications. These results quantitatively expose the limitations imposed by the tensor-network approximation inherent to DMRG in the dynamic-correlation regime.

Benchmark studies of the type presented here provide a rigorous and systematic means of assessing the performance of electronic structure methods across diverse correlation regimes and molecular symmetries. Such assessments enable the delineation of the domains of validity of approximate methods, establishing confidence in their predictive power for electronic structure properties (\emph{e.g.}, bonding, covalency, and polarization), spectroscopic observables (UV/Vis, X-ray, and related probes), and reaction energetics and pathways.

\section*{Supporting Information} Total energies and number of wavefunction parameters for the systems treated in this work.

\section{Acknowledgment}
The development of variational relativistic multi-reference methods is supported by the U.S. Department of Energy, Office of Science, Basic Energy Sciences, in the Heavy-Element Chemistry program (Grant No. DE-SC0021100).
The development of the Dirac--Coulomb--Breit method is supported by the U.S. Department of Energy, Office of Science, Basic Energy Sciences, in the Computational and Theoretical Chemistry program (Grant No. DE-SC0006863). The development of the Chronus Quantum computational software is supported by the Office of Advanced Cyberinfrastructure, National Science Foundation (Grants No. OAC-2103717 and OAC-2103705). SHY acknowledges support from the Florida State University Quantum Initiative. 

\newpage 

\begin{tocentry}
\centering
\includegraphics[width=3.3in]{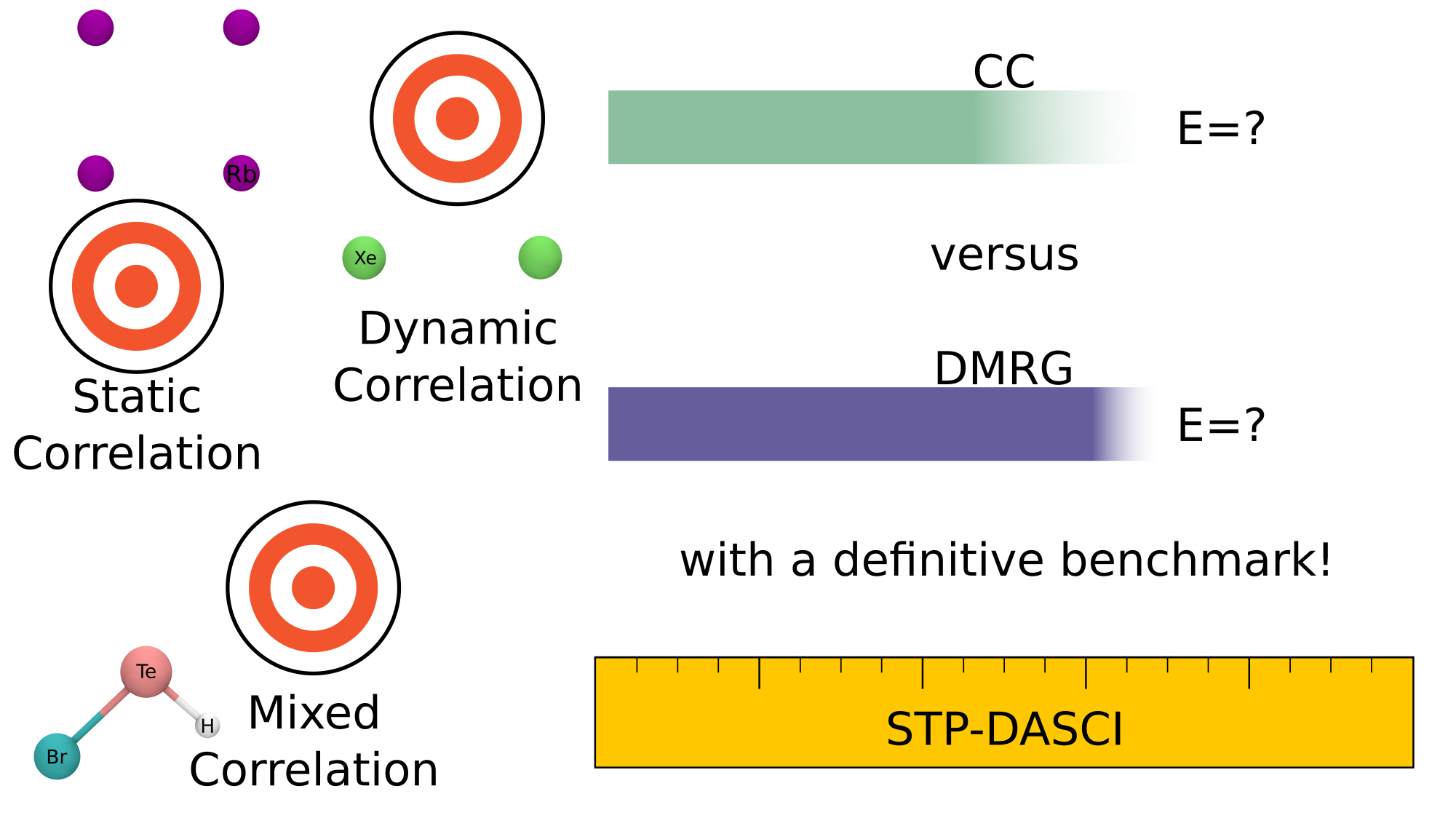}
\end{tocentry}


\bibliography{
    references/Journal_Short_Name,
    references/Li_Group_References, 
    references/CI,
    references/DMRG,
    references/sparseDAS,
    references/mcscf,
    references/exp_comparison,
    references/rel_other,
    references/CI_approx,
    references/RelCC
} 

\end{document}